\documentclass[showpacs,aps,fleqn,twocolumn]{revtex4}
\usepackage{graphicx}
\usepackage{amsmath}

\begin{document}

 \title{  Device-independent quantum secure direct communication }
\author{Lan Zhou$^{1}$,  Yu-Bo Sheng$^{2}$,\footnote{shengyb@njupt.edu.cn} and Gui-Lu Long$^{3,4}$\footnote{gllong@tsinghua.edu.cn}}
\address{
$^1$ School of Science, Nanjing University of Posts and Telecommunications, Nanjing, 210003, China\\
$^2$Institute of Quantum Information and Technology, Nanjing University of Posts and Telecommunications, Nanjing, 210003, China\\
$^3$State Key Laboratory of Low-dimensional Quantum Physics, Beijing, 100084, China\\
$^4$Department of Physics, Tsinghua University, Beijing, 100084, China}

\begin{abstract}
"Device-independent" not only represents a relaxation of the security assumptions about the internal
working of the quantum devices, but also can enhance the security of the quantum communication. In the paper, we put forward the first device-independent quantum secure direct communication (DI-QSDC) protocol, where no assumptions are made about the way the devices work or on what quantum system they operate. We show that in the absence of noise, the DI-QSDC protocol is absolutely secure and there is no limitation for the communication distance. However, under practical noisy quantum channel condition, the photon transmission loss and photon state decoherence would reduce the communication quality and threaten its absolute security. For solving the photon transmission loss and decoherence problems, we adopt noiseless linear amplification (NLA) protocol and entanglement purification protocol (EPP) to modify the DI-QSDC protocol. With the help of the NLA and EPP, we can guarantee the absolute security of the DI-QSDC and effectively improve its communication quality.
\end{abstract}

\pacs{03.67.Dd, 42.50.Dv, 42.50.Ex} \maketitle

\section{Introduction}
Quantum cryptography  can provide an absolute approach to guarantee the security of communication based on the basic principles of quantum mechanics. Quantum cryptography began
with quantum key distribution (QKD), i. e., BB84 protocol \cite{BB84}. Besides QKD, quantum cryptography includes some other branches, such as  quantum secret sharing \cite{QSS1} and quantum secure direct communication (QSDC) \cite{QSDC1}.
 QKD can share a series of secure keys between the
sender (Alice) and the receiver (Bob) \cite{BB84,QKD1,QKD2,QKD3,QKD4}. In QKD, in order to realize the secure communication, the sender and the receiver should ensure that the encryption and decrypt processes are absolutely secure. Moreover, they also require one-time pad and perfect key management.
Different from QKD, QSDC provides us another secure communication approach. QSDC allows the sender to transmit secret messages to the receiver without sharing a key first. QSDC has no key, no ciphertext, and either no key management.

The first QSDC protocol was proposed by Long \emph{et al.}, which exploited the properties of
entanglement and a block transmission technique \cite{QSDC1}.  QSDC can also be used to achieve QKD with high capacity \cite{RMP}.
 In 2003, the standard of QSDC was proposed \cite{QSDC2}. Later, QSDC based on single photon and high dimension  system were proposed \cite{QSDC3,QSDC4}.
  In the aspect of experiment, in 2016, Hu \emph{et al.}  experimentally realized the QSDC with single photons in a noisy
environment using frequency coding \cite{QSDCexperiment1}. In 2017, Zhang \emph{et al.} successfully completed the QSDC experiment with quantum memory \cite{shengprl}. Recently,  Zhu \emph{et al.} realized the first long-distance QSDC experiment in fibre \cite{experiment2}.

As a quantum cryptography mode, QSDC is also required to have absolute security. In QSDC protocols, photons should be transmitted in quantum channel for two rounds. For ensuring its absolute security and correctness, a security checking should be performed after each photon transmission process \cite{QSDC1,QSDC2,QSDC3}. The security checking method is often similar as that of QKD protocols, such as the BB84 protocol. The security analysis for QKD has been widely discussed \cite{security1,security2,security3,security4,security5,Lo}. They often assume that Alice and Bob have (almost) perfect control of the state preparation and the measurement devices. However, in practical experiment, this assumption is often critical. On the other hand, it has been proved that the absolute security of BB84 protocol relies on the assumption that Alice's and Bob's measurements act on a two-dimensional space. If we relax this assumption to the device-independent scenario, where Alice's and Bob's measurements are extended to act on a four-dimensional space, the security of BB84 protocol is no longer
guaranteed \cite{DI1,DI2,DIQKD3}.  In 2006, Ac\'{\i}n \emph{et al.} firstly proposed the device-independent quantum key distribution (DI-QKD) protocol \cite{DIQKD1}, which represents a relaxation of the security assumptions made in traditional QKD protocols \cite{DIQKD2,DIQKD3,DIQKD4}. It only requires that
quantum physics is correct and Alice and Bob do not allow any unwanted signal to escape from their
laboratories. DI-QKD requires the Alice and Bob to perform the Bell test \cite{Bell} (typically, the Clauser-Horne-Shimony-Holt (CHSH) test \cite{CHSH1}) on the pairs of entangled quantum systems shared between them. The violation of a Bell (CHSH) inequality is a necessary requirement for
the security of a QKD protocol in the general device-independent scenario.

Inspired by DI-QKD protocols \cite{DI2,DIQKD1,DIQKD3,DIQKD2}, in the paper, we propose the first DI-QSDC protocol, which can lower the requirement of experimental devices and enhance its security in device-independent scenario. In the protocol, we treat the quantum apparatuses as black boxes that produce classical outputs, possibly depending on the value of some classical inputs. We perform the CHSH tests in both two security checking processes and use the violation of CHSH inequality to guarantee the security of the DI-QSDC protocol. Similar as DI-QKD protocols, the DI-QSDC protocol is absolutely secure under ideal quantum channel condition. However, in practical noisy quantum channel condition, the photon transmission loss and photon state decoherence caused by the environmental noise seriously influence the DI-QSDC. First, they weaken the non-local correlation between Alice's and Bob's measurement results so that the maximal secure communication distance of the DI-QSDC protocol is largely limited. Second, even within the tolerate communication distance, the eavesdropper still has an opportunity to steal some secret messages without being detected, for there is in principle no way of distinguishing the entanglement
with an eavesdropper (caused by her measurements) from the entanglement with the environment (caused by innocent
noise). Third, they may unavoidably lead to message loss and message error. Therefore, in order to perform absolutely secure and practical DI-QSDC protocol in practical noisy quantum channel condition, we must solve the transmission photon loss and decoherence problems. Fortunately, the transmission photon loss problem can be solved by the noiseless linear amplification (NLA) \cite{TCR,NLA5,NLA6,NLA7,NLA8,NLA9}. Recently, the NLA protocols have been used in DI-QKD protocols to compensate the photon transmission loss, which can effectively increase the key rate and the maximal secure communication distance \cite{NLA1,NLA10,NLAadd,NLA11}. Meanwhile, the decoherence problems can be effectively solved by entanglement purification protocol (EPP) \cite{purification,EPP1,EPP2,EPP3,EPP4}. In this way, we adopt the NLA and EPP to modify our DI-QSDC protocol. With the help of NLA and EPP, we can increase the secure communication distance and guarantee the absolute security of our DI-QSDC protocol. Meanwhile, we can eliminate the message loss and effectively reduce the quantum bit error. Therefore, our modified DI-QSDC protocol has application potential in practical noisy quantum channel condition.

The paper is organized as follows. In Sec. 2, we explain our DI-QSDC protocol. In Sec. 3, we analyze the security and communication quality of the DI-QSDC protocol.  In Sec. 4, we improve the DI-QSDC protocol with NLA and EPP to resist to practical channel noise. In Sec. 5, we analyze the security and communication quality of the modified DI-QSDC protocol. In Sec. 6, we make a conclusion.

\section{The DI-QSDC protocol}
\begin{figure*}
\centering
\includegraphics[width=14cm,angle=0]{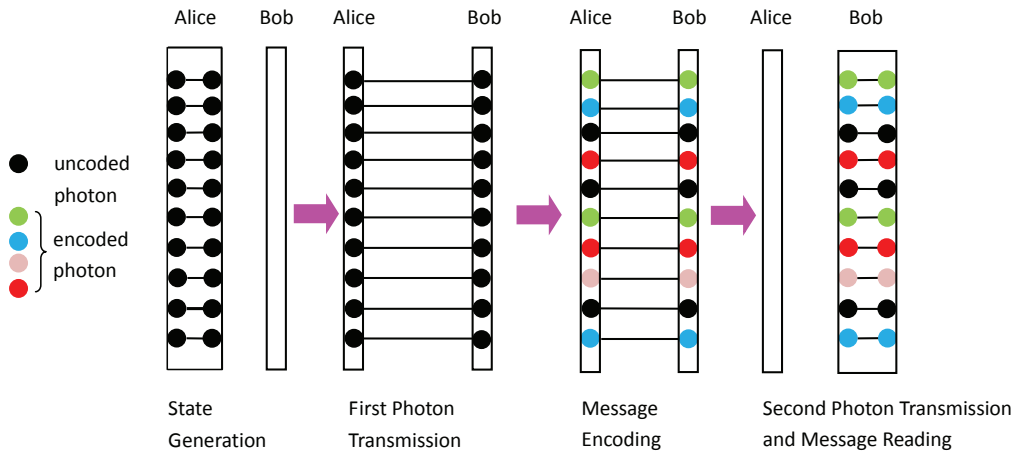}
\caption{The schematic principle of our DI-QSDC protocol. In the protocol, Alice prepares the ordered $N$ EPR pairs in $|\phi^{+}\rangle$. She divides the $N$ EPR pairs into two sequences and sends one photon sequence to Bob. If the first photon transmission process is secure, Alice encodes the EPR pairs using four unitary
operations and sends the other photon sequence to Bob. Finally, Bob can read the secret messages from Alice with the help of Bell-state analysis. In both two photon transmission processes, we randomly select a sufficiently large subset of EPR pairs to make the security checking. We introduce the "device-independent" mode in both two security checking rounds, which can guarantee the absolute security of our protocol.}
\end{figure*}

 Our DI-QSDC protocol shows that its security is based on a minimal set of fundamental assumptions. First, no unwanted information from Alice's and Bob's physical locations can leak to outside. Second, quantum physics is correct. Meanwhile, as we mainly analyze the effect from the practical noisy quantum channel, we assume the perfect entanglement source which can generate maximally entangled Bell states, the perfect quantum memory device with the storage efficiency of 100\%, and the lossless photon detector with the detection efficiency of 100\%. The basic principle of our DI-QSDC protocol is shown in Fig. 1. Before operating the DI-QSDC protocol, Alice and Bob have reached a consensus that each of the four Bell states carries two-qubit of classical information, say 00, 01, 10, and 11 for $|\phi^{+}\rangle$, $|\phi^{-}\rangle$, $|\psi^{+}\rangle$, and $|\psi^{-}\rangle$, respectively. $|\phi^{\pm}\rangle$ and $|\psi^{\pm}\rangle$ are four maximally entangled Bell states, which can be written as
\begin{eqnarray}
|\phi^{\pm}\rangle&=&\frac{1}{\sqrt{2}}(|HH\rangle\pm|VV\rangle),\nonumber\\
|\psi^{\pm}\rangle&=&\frac{1}{\sqrt{2}}(|HV\rangle\pm|VH\rangle).\label{bell}
\end{eqnarray}
 Here, $|H\rangle$ and $|V\rangle$ mean the horizontal and vertical polarization of the photon qubit, respectively.

The DI-QSDC protocol can be described as follows.

Step 1: Alice prepares an ordered $N$ EPR pairs in the state of $|\phi^{+}\rangle$ in her laboratory ($N$ is large). She divides the $N$ EPR pairs into two photon sequences, including the checking ($C$) photon sequence [$C_{1}$, $C_{2}$, $C_{3}$, $\cdots$, $C_{N}$] and the message ($M$) photon sequence [$M_{1}$, $M_{2}$, $M_{3}$, $\cdots$, $M_{N}$]. Then, she sends the photons in the $C$ sequence to Bob successively through the quantum channel.

Step 2: Due to the environmental noise, the single photon may be completely lost during the transmission process. The transmission efficiency $\eta$ is a function of the communication distance $d$, which can be written as \cite{loss}
\begin{eqnarray}
\eta=10^{-\alpha d/10}.\label{dd}
\end{eqnarray}
   The value of $\alpha$ is minimal in the "telecom window" around 1550 $nm$ ($\alpha=0.2$ $dB/km$). On the other hand, the environmental noise may also cause photon state decoherence. We assume the decoherence makes $|\phi^{+}\rangle$ degrade to a two-photon Werner state $\rho_{AB}$ with the form of \cite{DIQKD3}
\begin{eqnarray}
 \rho_{AB}=p|\phi^{+}\rangle_{AB}\langle \phi^{+}|+(1-p)\frac{I}{4}.\label{white}
 \end{eqnarray}
 Considering both photon transmission loss and decoherence, Alice and Bob finally share $N$ mixed states as
\begin{eqnarray}
\rho_{1}&=&\eta\rho_{AB}+\frac{1}{2}(1-\eta)(|H\rangle_{A}\langle H|+|V\rangle_{A}\langle V|)\nonumber\\
&=&\eta p|\phi^{+}\rangle_{AB}\langle \phi^{+}|+\eta(1-p)\frac{I}{4}\nonumber\\
&+&\frac{1}{2}(1-\eta)(|H\rangle_{A}\langle H|+|V\rangle_{A}\langle V|).\label{mixed}
\end{eqnarray}

   For ensuring the security of the first photon transmission process, Alice and Bob make the first round of security checking. In detail, Alice first randomly selects a large sufficiently subset of photons from the $M$ photon sequence to make the security checking and publics their positions to Bob through the classical channel. Then, they store others photons in the quantum memory device. Alice and Bob independently performs measurement chosen randomly on each of the security checking photons in the $M$ sequence and $C$ sequence, respectively. Alice has three possible measurements in the basis $|0\rangle\pm e^{iA_{i}}|1\rangle$ $(i=0,1,2)$, where $A_{0}=\frac{\pi}{4}$, $A_{1}=0$, and $A_{2}=\frac{\pi}{2}$. Bob has two possible measurements choices in the basis $|0\rangle\pm e^{iB_{i}}|1\rangle$ $(i=1,2)$, where $B_{1}=\frac{\pi}{4}$ and $B_{2}=-\frac{\pi}{4}$. All measurement results $a=\{a_{0}, a_{1}, a_{2}\}$ and $b=\{b_{1}, b_{2}\}$ have binary outcomes labeled by $a_{0}, a_{1}, a_{2}, b_{1}$, $b_{2}\in\{+1,-1\}$. If the parties obtain the inconclusive result (the photon detectors click no photon), the measurement result is set to be "+1" or "-1" randomly. After all the checking photon pairs have been measured, Alice and Bob reveal their measurement basis and measurement results in public classical channel. Then, they can estimate the CHSH polynomial as
\begin{eqnarray}
S_{1}=\langle a_{1}b_{1}\rangle+\langle a_{1}b_{2}\rangle+\langle a_{2}b_{1}\rangle-\langle a_{2}b_{2}\rangle,\label{S1}
\end{eqnarray}
 where $\langle a_{i}b_{j}\rangle$ is defined as $P(a=b|ij)-P(a\neq b|ij)$. Without loss of generality, we suppose that the marginal of all the measurements are random, such as $\langle a_{i}\rangle=\langle b_{j}\rangle=0$, $i\in\{0,1,2\}$ and $j\in\{1,2\}$.

 The amount of correlations between Alice's and Bob's symbols is also quantified by the quantum bit error rate (QBER) defined as
\begin{eqnarray}
Q_{1}=P(a_{0}\neq b_{1}).\label{q1}
 \end{eqnarray}
 According to the description above, both the photon transmission loss and decoherence would increase the value of QBER and reduce the CHSH polynomial. In theory, $Q_{1}$ can be calculated as
\begin{eqnarray}
Q_{1}=\frac{1}{2}(1-\eta)+\eta(\frac{1}{2}-\frac{p}{2})=\frac{1}{2}(1-\eta p).\label{error1}
 \end{eqnarray}
 As shown in Refs. \cite{DIQKD1,DIQKD2,DIQKD3}, the theoretical value of CHSH polynomial corresponding to $\rho_{1}$ in Eq. (\ref{mixed}) can be written as
\begin{eqnarray}
 S_{1}=2\sqrt{2}p\eta=2\sqrt{2}(1-2Q_{1}).\label{S11}
\end{eqnarray}

In practical experiment, if $S_{1}\leq 2$ (the well-known CHSH inequality), it indicates that the measurement results from Alice and Bob are classically correlated. Under this case, the photon transmission process is not secure. There exists a trivial attack for the eavesdropper that gives him complete information without being detected. In this way, if $S_{1}\leq 2$, Alice and Bob have to discard the whole DI-QSDC protocol. On the other hand, if the CHSH polynomial $2< S_{1}\leq 2\sqrt{2}$, it indicates the measurement results from Alice and Bob are quantum non-local correlated. Under this scale, we can bound the eavesdropper's photon interception rate. The maximal value $S_{1}=2\sqrt{2}$ corresponds to the case that Alice and Bob share maximally entangled states $|\phi^{+}\rangle_{AB}$ ($\eta=p=1$). In this case, the eavesdropper cannot intercept any photon in the first photon transmission process without being detected.

 Step 3: When the practical $S_{1}$ meets $2< S_{1}\leq 2\sqrt{2}$, Alice extracts the stored photons from the memory device. Then, Alice randomly selects a sufficiently large subset  of photons to be the checking photons and does not perform any operation on them. For the left photons, she encodes her  messages on them by performing one of the four unitary operations $U_{0}$, $U_{1}$, $U_{2}$, and $U_{3}$ on each of them. The four unitary operations can be written as
 \begin{eqnarray}
U_{0}&=&I=|H\rangle\langle H|+|V\rangle\langle V|,\nonumber\\
U_{1}&=&\sigma_{z}=|H\rangle\langle H|-|V\rangle\langle V|,\nonumber\\
U_{2}&=&\sigma_{x}=|V\rangle\langle H|+|H\rangle\langle V|,\nonumber\\
U_{3}&=&i\sigma_{x}=|H\rangle\langle V|-|V\rangle\langle H|.\label{encoding}
\end{eqnarray}
 They can make $|\phi^{+}\rangle$ evolve to $|\phi^{+}\rangle$, $|\phi^{-}\rangle$, $|\psi^{+}\rangle$, and $|\psi^{-}\rangle$, respectively, so that Alice can encode her secure messages 00, 01, 10 and 11 on the EPR pairs.
 It is worth noting that for preventing the eavesdropper to precisely intercept the corresponding $M$ photons during the second photon transmission process according to  his intercepted $C$ photons in the first photon transmission process, Alice messes up the $M$ photon sequence and records the corresponding position of each photon in the original $M$ sequence.

Step 4: Alice successively sends all the extracted photons in the $M$ sequence to Bob. After the photon transmission, Alice publics the position of each photon in the original $M$ sequence and the positions of the security checking photons by an authorized classical channel. Bob firstly recovers the $M$ photon sequence based on Alice's records  and store them into the quantum memory device. Then, he extracts his stored photons from the memory device according to the positions of security checking photons and makes the second security checking on the security checking photon pairs by his own.
In detail, Bob independently performs measurements on the two photons in each checking EPR pair randomly on the basis of \{$A_{0}$, $A_{1}$, $A_{2}$\} and \{$B_{1}$, $B_{2}$\}, respectively. After the measurements of all the checking EPR pairs, he can estimate the CHSH polynomial ($S_{2}$) and QBER ($Q_{2}$) after the second photon transmission process. Similarly, the photon transmission loss and decoherence also obviously influence $S_{2}$ and $Q_{2}$.
In theory, after the second photon transmission, the security checking photon state in Bob's hand has the form of
\begin{eqnarray}
\rho_{2}&=&\eta^{2}\rho'_{AB}+\frac{\eta(1-\eta)}{2}(|H\rangle_{b}\langle H|+|V\rangle_{b}\langle V|)\nonumber\\
&+&\frac{\eta(1-\eta)}{2}(|H\rangle_{a}\langle H|+|V\rangle_{a}\langle V|)\nonumber\\
&+&(1-\eta)^{2}|vac\rangle\langle vac|,
\end{eqnarray}
where
\begin{eqnarray}
\rho'_{AB}=p^{2}|\phi^{+}\rangle\langle \phi^{+}|+\frac{(1-p^{2})}{4}I.
\end{eqnarray}
As a result, the theoretical value of $Q_{2}$ and $S_{2}$ can be calculated as
\begin{eqnarray}
Q_{2}&=&\frac{1}{2}(1-\eta^{2}p^{2}),\\\label{error2}
S_{2}&=&2\sqrt{2}p^{2}\eta^{2}=2\sqrt{2}(1-Q_{2}).\label{S3}
\end{eqnarray}
It can be found that $Q_{2}$ is the total QBER after two rounds of photon transmission. It is obvious that $Q_{2}>Q_{1}$ and $S_{2}<S_{1}$.

For bounding eavesdropper's photon interception rate in the second photon transmission process, we also require that $2<S_{2}\leq2\sqrt{2}$. Otherwise, Alice and Bob should discard the whole communication.

Step 5: If the parties entrust both the two photon transmission processes, Bob can finally read out the messages from Alice by performing Bell-state analysis on the message encoded EPR pairs in his hand \cite{analysis1,analysis2,analysis3}.

\section{Security and communication quality of the DI-QSDC protocol}
 In the security analysis, we only require the eavesdropper to obey the laws of quantum physics, and no other limitations are imposed to him.
 As description in above section, here we assume the perfect entanglement source, the perfect quantum memory device, and the lossless photon detector, and analyze the effect on DI-QSDC from quantum channel noise. We first analyze the security and communication quality of our DI-QSDC protocol under perfect quantum channel condition ($\eta=p=1$). Under this case, the CHSH polynomial $S_{1}$ and $S_{2}$ equal to $2\sqrt{2}$, and the QBERs in both two photon transmission processes are exactly 0. As a result, the eavesdropper cannot intercept any photon in both two photon transmission processes without being detected, so that our DI-QSDC protocol is absolutely secure. We define the communication efficiency ($E_{c}$) is the amount of transmitted correct secure information qubits divided by the total amount of the information qubits. Under perfect quantum channel condition, it is obvious that $E_{c}$ can reach the maximum of 1.

On the other hand, under the practical noisy quantum channel condition, the situation is different. The security of our DI-QSDC protocol relies on
the CHSH polynomials $S_{1}$ and $S_{2}$ violating the CHSH inequality. The photon transmission efficiency $\eta$ and the value of $p$ both affect the values of $S_{1}$ and $S_{2}$. Considering $\eta$ decreases with the growth of communication distance, the secure communication distance of DI-QSDC protocol will be limited.  Meanwhile, according to the step 3 and step 4 of the DI-QSDC protocol, before the second photon transmission process, Alice would mess up the $M$ photon sequence. This performance can effectively avoid the precise interception of the eavesdropper in the second photon transmission process.  As a result, under the scale of $2<S_{1},S_{2}\leq2\sqrt{2}$,  the eavesdropper can only intercept some photons randomly in both two photon transmission process. Based on the estimation of DI-QKD protocol in Ref. \cite{DIQKD3,NLA1}, we can respectively bound the photon interception rate ($I$) of the eavesdropper in the two photon transmission processes by
\begin{eqnarray}
I_{1}(S_{1})&\leq& \chi(S_{1})=h(\frac{1+\sqrt{(S_{1}/2)^{2}-1}}{2}),\nonumber\\
I_{2}(S_{2})&\leq& \chi(S_{2})=h(\frac{1+\sqrt{(S_{2}/2)^{2}-1}}{2}).
\end{eqnarray}
Here, $h$ is the binary entropy with the form of
\begin{eqnarray}
h(x)=-xlog_{2}(x)-(1-x)log_{2}(1-x).
\end{eqnarray}
 As $S_{2}<S_{1}$, it is obvious that the photon interception rate in the second photon transmission process ($I_{2}$) is higher than that ($I_{1}$) in the first photon transmission process. Only if the eavesdropper can intercept both the two photons of an EPR pair in the two photon transmission processes, he can finally read out the 2 bits of message encoded in the EPR pair by the Bell-state analysis. Therefore, we can bound the information interception rate ($I_{E}$) of the eavesdropper by
\begin{eqnarray}
I_{E}\leq I_{1}(S_{1}).\label{St}
\end{eqnarray}
If the eavesdropper only intercepts one photon in the second photon transmission process but does not intercept the corresponding photon in the first photon transmission process, he can not read out the encoded information. However, under this case, the eavesdropper's interception could disturb the communication, which may makes Bob read out the incorrect information.

Similarly, based on the estimation of DI-QKD protocol in Ref. \cite{DIQKD3,NLA1}, the achievable communication efficiency ($E_{c1}$) of the DI-QSDC protocol can be given by
\begin{eqnarray}
E_{c1}\geq1-h(Q_{2})-I_{2}(S_{2}).\label{E1}
\end{eqnarray}

\begin{figure}[!h]
\begin{center}
\includegraphics[width=7cm,angle=0]{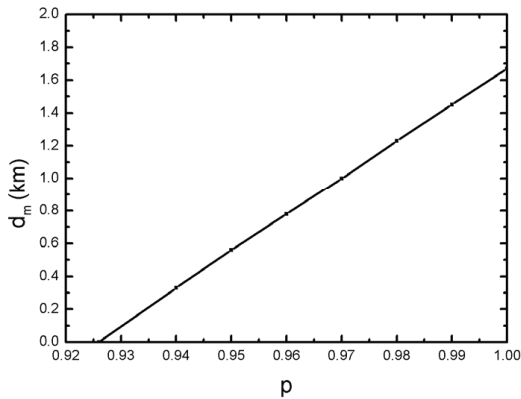}
\caption{The maximal communication distance $d_{m}$ is plotted as a function of $p$ in the device-independent scenario.  }
\end{center}
\end{figure}

\begin{figure}[!h]
\begin{center}
\includegraphics[width=7cm,angle=0]{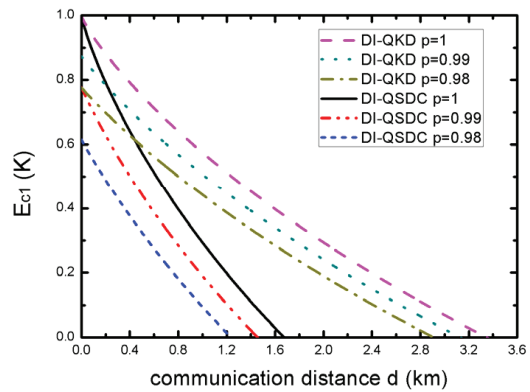}
\caption{The key generation rate ($K$) of the DI-QKD protocol in Ref. \cite{DIQKD3} and $E_{c1}$ of the DI-QSDC protocol as a function of the communication distance $d$ in the device-independent scenario. Here, in both DI-QKD and DI-QSDC protocols, we control the value of $p$ to be 1, 0.99, and 0.98, respectively.    }
\end{center}
\end{figure}

From Eq. (\ref{E1}), we define the maximal communication distance $d_{m}$ of the DI-QSDC protocol corresponding to $E_{c1}=0$. It can be calculated that for obtaining $E_{c1}\geq0$,  the threshold value of $p$ must be as high as $0.926$. In Fig. 2, we show the values of $d_{m}$ as a function of $p$. It can be found that when $p=1$, which means the decoherence in the transmission process can be neglected, we can obtain the maximal communication distance of the DI-QSDC protocol is only 1.67 $km$. With the decrease of $p$, the value of $d_{m}$ decrease obviously. When $p$ decreases to the threshold of $0.926$, $d_{m}=0$. It means when the decoherence makes $p<0.926$, no correct information can be transmitted from Alice to Bob.

In Fig. 3, we compare the communication efficiency of our DI-QSDC protocol with the key generation rate ($K$) of the DI-QKD protocol in Ref. \cite{DIQKD3}. Here, we control the value of $p$ to be 1, 0.99, and 0.98, respectively. It can be found that with the decrease of $p$, both the $K$ of DI-QKD protocol and  $E_{c1}$ of DI-QSDC protocol reduces. Meanwhile, $E_{c1}$ of our DI-QSDC protocol is smaller than $K$ of DI-QKD protocol ($K\geq1-h(Q_{1})-I_{1}(S_{1})$), and its maximal communication distance (1.67 km) is also lower than that of the DI-QKD (about 3.4 $km$). The reason is that the DI-QSDC requires two rounds of photon transmission while DI-QKD requires only one. As a result, the influences from the environmental noise and interception on DI-QSDC protocol are more serious than that on the DI-QKD.

 On the other hand, it can be found that the information received by Bob may be incomplete and partly incorrect. We define the information loss rate ($r_{loss}$) as the amount of lost information qubits divided by the total amount of the information qubits, and the information error rate ($r_{error}$) as the amount of incorrect qubits read out by Bob divided by the total amount of the information qubits that Bob can read out. We can calculate $r_{loss}$ and $r_{error}$ as
 \begin{eqnarray}
 r_{loss}&=&1-\eta^{2},\\
 r_{error}&=&1-\frac{1+3p^{2}}{4}=\frac{3}{4}(1-p^{2}).\label{rerror}
 \end{eqnarray}

\section{Modification of our DI-QSDC protocol resist practical channel noise }
 According to above security analysis, the photon transmission loss and decoherence caused by the channel noise threaten the absolute security of the DI-QSDC and seriously limit its communication efficiency and communication distance. For overcoming the photon transmission loss and decoherence problem during the photon transmission process, we adopt the NLA protocol and EPP  to modify the DI-QSDC protocol.

\subsection{Noiseless linear quantum amplification protocol}
We introduce a deterministic entanglement-based NLA protocol from Ref. \cite{NLA7} to solve the photon transmission loss problem. The basic principle of this NLA protocol is shown in Fig. 4. Suppose  Alice generates an EPR pair in the state of $|\phi^{+}\rangle$ and sends one photon in the EPR pair to a distant party Bob. After transmitting through a noisy quantum channel with the transmission efficiency of $\eta$, the target state $|\phi^{+}\rangle_{a_{0}b_{0}}$ between Alice and Bob degrades to a mixed state as
 \begin{eqnarray}
\rho_{in}=\eta|\phi^{+}\rangle_{a_{0}b_{0}}\langle\phi^{+}|+\frac{1}{2}(1-\eta)(|H\rangle_{a_{0}}\langle H|+|V\rangle_{a_{0}}\langle V|).\nonumber\\
\end{eqnarray}

\begin{figure}[!h]
\begin{center}
\includegraphics[width=8cm,angle=0]{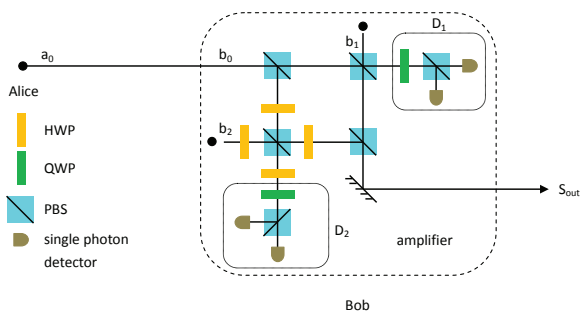}
\caption{Basic setup of the linear-optical heralded noiseless amplification (NLA) protocol for the EPR pair. PBS is a polarizing beam splitter. $D_{1}$ and $D_{2}$ are standard polarization analysis blocks shown in Ref. \cite{D}. HWP represents a half-wave plate and QWP represents a quarter-wave plate. Bob selects the items which make $D_{1}$ and $D_{2}$ each detect exactly one photon. In detail, Bob performs a projection polarization measurement in the output ancillary ports with four single photon detectors. When the detection result is $|HH\rangle_{D_{1}D_{2}}$ or $|VV\rangle_{D_{1}D_{2}}$, he can finally distill the target state $|\phi^{+}\rangle_{a_{0}S_{out}}$, while when the detection result is $|HV\rangle_{D_{1}D_{2}}$ or $|VH\rangle_{D_{1}D_{2}}$, he can distill $|\phi^{-}\rangle_{a_{0}S_{out}}$, which can be easily converted to $|\phi^{+}\rangle_{a_{0}S_{out}}$ with the feed-forward correction.}
\end{center}
\end{figure}

For overcoming the photon transmission loss, Bob should prepare two ancillary photons in a maximally entangled Bell state as
 \begin{eqnarray}
|\phi^{+}\rangle_{b_{1}b_{2}}=\frac{1}{\sqrt{2}}(|HH\rangle_{b_{1}b_{2}}+|VV\rangle_{b_{1}b_{2}}).
\end{eqnarray}
Bob adopts four half-wave plates (HWPs) to perform horizontal-vertical polarization swapping and six polarizing beam splitters to transmit the photons in $|H\rangle$ and reflect the photons in $|V\rangle$, respectively. After performing the amplification, Bob can only distill the items $|H\rangle_{a_{0}}|HHH\rangle_{b_{0}b_{1}b_{2}}$ and $|V\rangle_{a_{0}}|VVV\rangle_{b_{0}b_{1}b_{2}}$ which make the standard polarization analysis blocks $D_{1}$ and $D_{2}$ each detect exactly one photon. Then, by making the single photon entering $D_{1}$ ($D_{2}$) pass through a quarter-wave plate (QWP) and performing the polarization measurements on the output photons, Bob can finally eliminate the vacuum state and distill the pure target state $|\phi^{+}\rangle_{a_{0}S_{out}}$ with the total success probability of
\begin{eqnarray}
P_{NLA}=0.5\eta.\label{PNLA}
\end{eqnarray}

Similarly, if the target state is one of the other three Bell-states in Eq. (\ref{bell}), this NLA protocol will also work. As a result, the NLA can completely eliminate the photon transmission loss problem, but cannot do anything about the decoherence.

\subsection{Entanglement purification protocol}
\begin{figure}[!h]
\begin{center}
\includegraphics[width=8cm,angle=0]{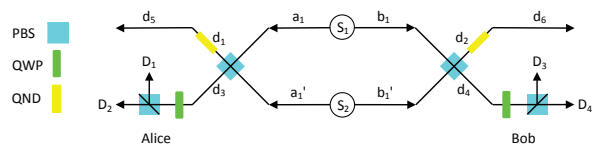}
\caption{The schematic principle of the linear-optical entanglement purification protocol (EPP). Alice and Bob share two pairs of two-photon Werner state in Eq. (\ref{white}). They superpose the photons on two PBSs, respectively and keep the items which make each of the four output modes $d_{1}d_{2}d_{3}d_{4}$ contain exactly one photon \cite{EPP2}. For preserving the distilled photon state, Alice and Bob adopt two linear-optical QND gates in $d_{1}d_{2}$ modes, respectively, whose structure is the same as the amplifier shown in Fig. 4. Then, Alice and Bob make the photons in the $d_{3}d_{4}$ modes pass through the QWP and perform the polarization measurements on each of the output photons. If the measurement result is $|HH\rangle_{D_{2}D_{4}}$ or $|VV\rangle_{D_{1}D_{3}}$, they can finally distill a new mixed state $\rho_{out1}=A_{1}|\phi^{+}\rangle_{d_{5}d_{6}}\langle\phi^{+}|+B_{1}|\psi^{+}\rangle_{d_{5}d_{6}}\langle\psi^{+}|
+C_{1}|\phi^{-}\rangle_{d_{5}d_{6}}\langle\phi^{-}|
+D_{1}|\psi^{-}\rangle_{d_{5}d_{6}}\langle\psi^{-}|$. If the measurement result is $|HV\rangle_{D_{2}D_{3}}$ or $|VH\rangle_{D_{1}D_{4}}$, they can finally distill  $\rho_{out2}=C_{1}|\phi^{+}\rangle_{d_{5}d_{6}}\langle\phi^{+}|+D_{1}|\psi^{+}\rangle_{d_{5}d_{6}}\langle\psi^{+}|
+A_{1}|\phi^{-}\rangle_{d_{5}d_{6}}\langle\phi^{-}|
+B_{1}|\psi^{-}\rangle_{d_{5}d_{6}}\langle\psi^{-}|$, which can be easily converted to $\rho_{out1}$ after Alice performing a local phase-flip operation on the photon in $d_{5}$. }
\end{center}
\end{figure}

  The schematic principle of the linear-optical EPP is shown in Fig. 5. We suppose that Alice and Bob use two pairs of $\rho_{AB}$ in Eq. (\ref{white}) in the $a_{1}b_{1}$ and $a_{1}'b_{1}'$ modes, respectively. As shown in Ref. \cite{EPP2}, they pass the photons in $a_{1}a_{1}'$ and $b_{1}b_{1}'$ through two PBSs, respectively and select the items corresponding to exactly one photon in each of the four output modes of the PBSs ("four-mode cases"). However, as all the photons are detected by the single photon detectors, the distilled new mixed state cannot be preserved for other applications.

  On the other hand, it can be found that the setup of the amplifier in Fig. 4 actually plays the role of a linear-optical quantum nondemolition detector (QND) gate, which can distinguish the input photon number 0 and 1 without destroying the single photons. According to Eq. (\ref{PNLA}), the success probability of each QND gate is $\frac{1}{2}$. In this way, we introduce two above QND gates in $d_{1}$ and $d_{2}$ modes to preserve the distilled photons. Then, Alice and Bob make the photons in the $d_{3}d_{4}$ modes pass through the QWPs and perform the polarization measurements on each of the output photons. Finally, they can distill a new mixed state in the $d_{5}d_{6}$ modes as
\begin{eqnarray}
\rho_{out1}&=&A_{1}|\phi^{+}\rangle_{d_{5}d_{6}}\langle\phi^{+}|+B_{1}|\psi^{+}\rangle_{d_{5}d_{6}}\langle\psi^{+}|\nonumber\\
&+&C_{1}|\phi^{-}\rangle_{d_{5}d_{6}}\langle\phi^{-}|
+D_{1}|\psi^{-}\rangle_{d_{5}d_{6}}\langle\psi^{-}|,\label{error1n}
\end{eqnarray}
with the total success probability as
 \begin{eqnarray}
P_{E_{1}}&=&\frac{1+p^{2}}{4}\times\frac{1}{4}=\frac{1+p^{2}}{16}.
\end{eqnarray}

The fidelity of the new mixed state can be written as
\begin{eqnarray}
A_{1}&=&\frac{5p^{2}+2p+1}{4(1+p^{2})}.
\end{eqnarray}
It has been proved that $A_{1}$ is higher than the initial fidelity $p+\frac{1-p}{4}=\frac{1+3p}{4}$, when the initial fidelity satisfies  $\frac{1+3p}{4}>\frac{1}{2}$, say, $p>\frac{1}{3}$. Moreover, when the protocol is successful, Alice and Bob can also choose two pairs of $\rho_{out1}$ and repeat the entanglement purification process to further increase the fidelity of $|\phi^{+}\rangle$.

\subsection{Application of NLA protocol and EPP in the DI-QSDC protocol}
\begin{figure*}
\centering
\includegraphics[width=14cm,angle=0]{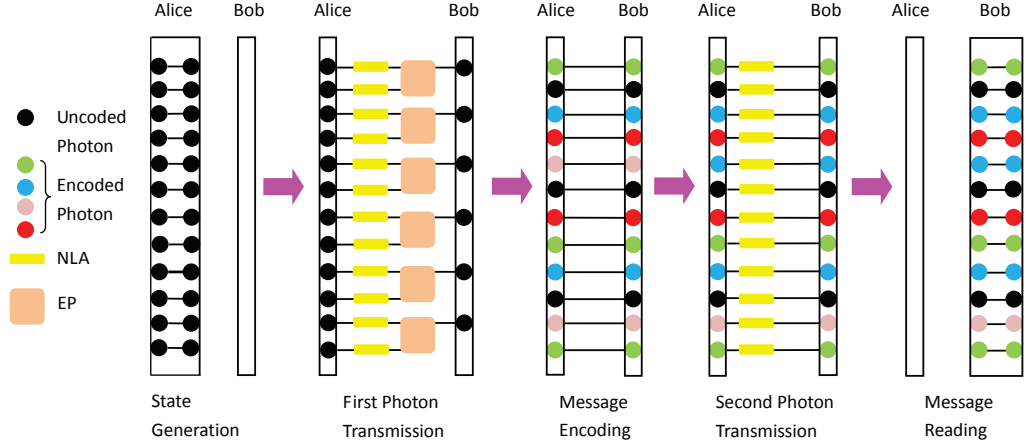}
\caption{The schematic principle of the modified DI-QSDC protocol. In the protocol, we adopt NLA and  EPP to solve the photon transmission loss and decoherence problems in the first photon transmission process and adopt NLA in the second photon transmission process to compensate the photon transmission loss. The modified DI-QSDC protocol can resist the practical channel noise and ensure its absolute security.}
\end{figure*}

Here, we adopt the NLA protocol and EPP in the DI-QSDC protocol to resist the practical channel noise. The schematic principle of our modified DI-QSDC protocol is shown in Fig. 6. In the first photon transmission process, we adopt the NLA protocol and EPP to construct the near-perfect entanglement channel.
In detail, Bob first uses the NLA protocol to compensate the transmission photon loss. Once Alice emits one photon, Bob performs the above NLA protocol. If the NLA protocol is successful, Bob will tell Alice by public classical channel to reserve the corresponding $M$ photon. Otherwise, if the NLA protocol fails, he will tell Alice to discard the corresponding $M$ photon. As a result, with the help of the NLA protocol, Alice and Bob can completely eliminate the photon transmission loss. Then, Alice and Bob repeat the EPP to solve the decoherence problem. The iteration number of EPP depends on the initial value of $p$ and the expected fidelity of $|\phi^{+}\rangle$ (i.e. 0.99). It is worth noting that the EPP works only when the initial fidelity satisfies $p>\frac{1}{3}$. If $p<\frac{1}{3}$, performing the EPP cannot increase the fidelity of $|\phi^{+}\rangle$. After NLA and EPP, Alice and Bob have successfully constructed the near-perfect entanglement channel.

   During the second photon transmission process, once Alice emits a single photon, Bob performs the NLA protocol. If the NLA protocol is successful,he tells Alice to encode her next message on the next photon¡£ If the NLA protocol fails, Bob discards the corresponding photon in the C sequence in his hand and tells Alice  to recodes this message on the next photon, and so on, until Bob successfully receives the encoded photon. However, the EPP cannot be used in the second photon transmission process, for it may change the encoded secret messages of the photons. As a result, the decoherence in the second photon transmission process is still unavoidable.

\section{Security and communication quality of the modified DI-QSDC protocol }
 With the help of the NLA and EPP in the first photon transmission process, $\eta$ can be increased to 1 and $p$ can be increased to be quite close to 1. After NLA and EPP, Alice and Bob can construct the near-perfect entanglement channel and the CHSH polynomial $S_{1}' \rightarrow 2\sqrt{2}$. Under this case, we can easily obtain $I_{1}'\rightarrow 0$ and $Q_{1}'\rightarrow 0$. Therefore, it is impossible for the eavesdropper to steal photons from the first photon transmission process without being detected and the absolute security of the first photon transmission process can be ensured. In the second photon transmission process, with the help of NLA, we can also obtain $\eta=1$. In this way, we can eliminate the message loss ($r'_{loss}=0$) and ensure the completeness of the transmitted message. However, as we cannot perform EPP on the encoded EPR pairs, the photon state decoherence in the second photon transmission process can reduce the value of CHSH polynomial and cause quantum bit error. The $S_{2}'$ and $Q_{2}'$ in the modified DI-QSDC protocol can be respectively rewritten as
\begin{eqnarray}
S_{2}'&=&2\sqrt{2}p,\\
 Q_{2}'&=&\frac{1}{2}-\frac{p}{2}.
 \end{eqnarray}
  $Q_{2}'$ is the total QBER in the modified DI-QSDC protocol. In this way, it is also possible for the eavesdropper to steal some photons from the second photon transmission process without being detected. In the device-independent scenario, we can bound the interception rate ($I_{2}'$) of the eavesdropper in the second photon transmission process by
\begin{eqnarray}
I_{2}'(S_{2}')&\leq& \chi(S_{2}')=h(\frac{1+\sqrt{(S_{2}'/2)^{2}-1}}{2}).
\end{eqnarray}
 According to Eq. (\ref{St}), as $I_{1}'\rightarrow 0$, we can obtain the eavesdropper's information interception rate of the modified DI-QSDC protocol as
\begin{eqnarray}
 I_{E}\rightarrow 0. \label{In}
 \end{eqnarray}
Eq. (\ref{In}) agrees to that fact that the security of the DI-QSDC can be ensured when we can ensure the absolute security of the first photon transmission process. The reason is that after two rounds of photon transmission, all the Bell-state analysis protocols require to perform
quantum operations or measurements on both the two photons of an EPR pair. As long as we can ensure the absolute security of the first photon transmission process, the eavesdropper
cannot read out any message even if he can steal some photons in the second photon transmission process. In this way, ensuring the absolute security of the first photon transmission process can guarantee the absolute security of the whole DI-QSDC.

Meanwhile, the decoherence in the second photon transmission process can also cause information error. The information error rate of the modified DI-QSDC can be written as
\begin{eqnarray}
r'_{error}=1-\frac{1+3p}{4}=\frac{3}{4}(1-p),
 \end{eqnarray}
which is lower than that of the original DI-QSDC protocol in Eq. (\ref{rerror}).

As the success probabilities of the NLA and EPP are lower than 1, the adoption of NLA and EPP would reduce the communication efficiency of the modified DI-QSDC protocol. In the first photon transmission process, we assume to repeat the EPP for $k$ times to increase the value of $p$. The communication efficiency $E_{c2}$ of the modified DI-QSDC can be written as
\begin{eqnarray}
E_{c2}\geq\frac{\eta^{4}P_{E_{1}}P_{E_{2}}\cdots P_{E_{K}}}{2^{k+2}}[1-h(Q_{2}')-I_{2}(S_{2}')].\ \label{e2}
\end{eqnarray}
For ensuring $E_{c2}\geq 0$, we can calculate the threshold value of $p$ to be 0.858, which is the same as that in the DI-QKD protocol in Ref. \cite{DIQKD3}. Comparing with the original DI-QSDC protocol, the adoption of EPP in the first transmission process reduces the threshold value of $p$ from 0.926 to 0.858. In this way, the modified DI-QSDC is easier to implement under practical noisy experimental condition.

On the other hand, the decoherence in the second photon transmission process can still cause message error. Comparing with original DI-QSDC protocol where the decoherence exists in both two photon transmission process, the message error in the modified DI-QSDC protocol decreases obviously.

\begin{figure}[!h]
\begin{center}
\includegraphics[width=8cm,angle=0]{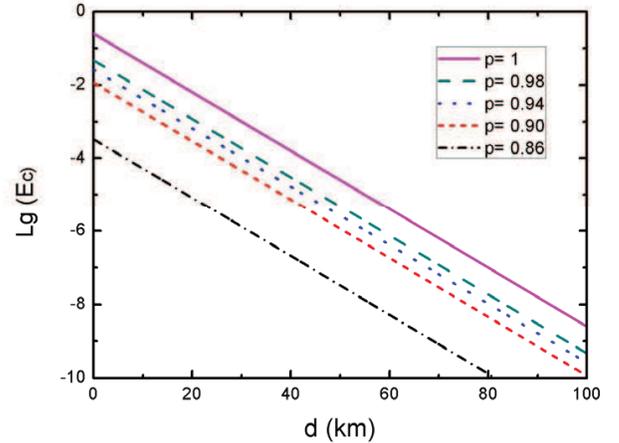}
\caption{The logarithmic communication efficiency ( Lg ($E_{c}$)) of the modified DI-QSDC protocol as a function of the practical communication distance ($d$) in the device-independent scenario. Here, we control the value of $p$ to be 1, 0.98, 0.94, 0.90, and 0.86, respectively, and set the target value of $p$ after the EPP to be above 0.99. }
\end{center}
\end{figure}

 In Fig. 7, we show the value of $Lg(E_{c2})$ as a function of the practical communication distance $d$ under the $p=1$, $0.98$, $0.94$, $0.90$, $0.86$, respectively. In the first photon transmission process, we aim to increase the fidelity of $|\phi^{+}\rangle\langle \phi^{+}|$ to be no less than 0.99 by repeating the EPP for $k$ times. In this way, when $p=1$, $0.98$, $0.94$, $0.90$, and $0.86$, we need to operate the EPP for $0$, $2$, $2$, $2$, and $3$ times, respectively. By performing both the NLA and EPP operations, we can ensure the absolute security of the modified DI-QSDC protocol and extend its maximal secure communication distance. The cost is that introducing NLA and EPP will lower the communication efficiency ($E_{c2}$) of the DI-QSDC. Although $E_{c2}$ is relatively low, in practical application, we can increase the initial input power to obtain suitable information amount. For example, we consider that the photon sources are excited with a repetition rate of 10 GHz \cite{frequency}. Without the NLA and EPP, the original DI-QSDC cannot transmit secure and correct information beyond 1.67 km even under $p=1$. While with the help of the NLA and EPP, the modified DI-QSDC can achieve communication rates of about 25 $bit/s$ under $p=1$ and about 1 $bit/s$ under $p=0.9$ on the distance of 100 km. Even when $p$ is very close to the threshold value, i.e., $p=0.86$, the modified DI-QSDC can still achieve the communication rate of about $1 bit/s$  on the distance of 80 km.

     The modified DI-QSDC protocol requires perfect entanglement source, which are necessary to guarantee its absolute security. On the other hand, the practical imperfect single-photon detectors and memory devices can also be used in the modified DI-QSDC protocol. Actually, the inefficiency of the single photon detector and memory device can be treated as photon transmission loss. The inefficiency of photon detector and the memory devices will reduce the communication efficiency of the modified DI-QSDC protocol. After ensuring the security, Bob can finally read out the secret messages by performing Bell-state analysis. In linear optics, all the Bell-state analysis protocols can only distinguish two of the four Bell states. As a result, the message reading efficiency of the DI-QSDC protocol is only 50\%. This will not influence the absolute security of the modified DI-QSDC protocol, but will make the practical message transmission efficiency be half of the communication efficiency. In practical experiment, we can only encode two secret messages 0 and 1 on the EPR pairs by performing two of the four unitary operations $U_{0}$, $U_{1}$, $U_{2}$, and $U_{3}$ on the $M$ photons. Under this case, the practical message transmission efficiency equals to the communication efficiency.

\section{Conclusion}
In the quantum secure communication field, DI not only represents a relaxation of the security assumptions made in usual quantum communication protocols, but also can enhance the security of the quantum communication, especially in the device-independent scenario. In the paper, we put forward the first DI-QSDC protocol. In the DI-QSDC protocol, Alice can directly send the secure message to Bob by two rounds of photon transmission. Similar as the previous DI-QKD protocols, the absolute security and correction of the DI-QSDC protocol can be ensured based on two rounds of non-locality test, where the CHSH inequality provides us an estimation of eavesdropper's knowledge and the communication efficiency. Here, we mainly consider the influence from the practical channel noise on the security and communication efficiency of DI-QSDC protocol.
In the absence of channel noise, the DI-QSDC protocol is absolutely secure and no quantum bit error exists. Meanwhile, there is no limit to its communication distance. However, under practical noisy  channel condition, the photon transmission loss and photon state decoherence have  serious influences on the DI-QSDC. The maximal communication distance of the DI-QSDC protocol is only 1.67 $km$, which is lower than that of the DI-QKD protocol (about 3.4 $km$). The threshold value of quantum state fidelity $p$ is 0.926. As the DI-QSDC protocol requires two photon transmission processes while DI-QKD only requires one, the eavesdropper can steal the encoded information only when he intercepts both the two photons in an EPR pair. In this way, the eavesdropper's information interception rate is lower than that of the DI-QKD. However, any quantum bit error and eavesdropper's photon interception in both two photon transmission process  can disturb the communication, which makes Bob cannot read out the correct information. As a result, the communication efficiency of DI-QSDC protocol is also lower than that of the DI-QKD protocol. Meanwhile, the photon transmission loss and decoherence would also lead to the information loss and information error, which will also seriously limit the practical application of DI-QSDC.

For overcoming the photon transmission loss and decoherence problems in the photon transmission process, we introduce the NLA protocol and EPP to modify the DI-QSDC protocol. We adopt the NLA protocol in both two photon transmission processes to completely compensate the photon transmission loss, so that, the information loss can be totally eliminated. We adopt the EPP in the first photon transmission process to construct the near-perfect quantum channel, so that the eavesdropper almost cannot steal any photon in the first photon transmission process without being detected. In the second photon transmission process, we cannot perform the EPP on the encoded photon EPR pairs, for the EPP may change the encoded secure information. As a result, the photon state decoherence still exists, which may cause the quantum bit error and provide the eavesdropper an opportunity to steal some photons without being detected. However, as the eavesdropper cannot perform Bell-state analysis when he only captures one of the two photons in an EPR pair, the absolute security of the first photon transmission process guarantees the absolute security of the whole modified DI-QSDC protocol. Meanwhile, due to the introduction of EPP in the first photon transmission process, the modified DI-QSDC protocol can reduce the threshold value of $p$ from 0.926 to 0.858 and reduce the information error. We assume that the photon sources are excited with a repetition rate of 10 GHz. The modified DI-QSDC can achieve communication rates of about 25 $bit/s$ under $p=1$ and about 1 $bit/s$ under $p=0.9$  on the distance of 100 $km$. Even when $p$ is very close to the threshold value ($p=0.86$), the modified DI-QSDC can still achieve the communication rate of about 1 $bit/s$  on the distance of 80 $km$. Based on above features, the modified DI-QSDC protocol may have application potential in the practical imperfect condition in the future.

\section*{ACKNOWLEDGEMENTS} This work is supported by the  National Natural Science Foundation
of China under Grant  No. 11474168, the China Postdoctoral Science Foundation under Grant No. 2018M642293, the open research
fund of the Key Lab of Broadband Wireless Communication
and Sensor Network Technology, Nanjing University of
Posts and Telecommunications, Ministry of Education, and a Project Funded by the Priority
Academic Program Development of Jiangsu Higher Education Institutions.

\end{document}